\documentclass{aa}
\usepackage{natbib}
\bibpunct{(}{)}{;}{a}{}{,}
\usepackage{graphicx}
\usepackage{txfonts}
\usepackage{xcolor}
\usepackage{placeins}
\usepackage{float}
\usepackage{lipsum}         
\usepackage[colorlinks=true, linkcolor=blue, citecolor=blue, urlcolor=blue]{hyperref}

\begin{document}
\title{The full evolution of the type-C QPO in MAXI J1348$-$630 revealed by \textit{Insight}-HXMT}
 
\author{
X.-L. Wang\inst{1,2}
\and Z. Yan\inst{2,3}\thanks{Corresponding author: \email{zyan@shao.ac.cn}}
\and F.-G. Xie\inst{2,3}
\and J.-F. Wang\inst{1,3}
\and Y.-X. Li\inst{2,4}
\and Z.-Y. Liu\inst{1}
\and R.-Y. Ma\inst{1,3}\thanks{Corresponding author: \email{ryma@xmu.edu.cn}}
}

\institute{
Department of Astronomy and Institute of Theoretical Physics and Astrophysics, Xiamen University, Xiamen, Fujian 361005, China
\and
Shanghai Astronomical Observatory, Chinese Academy of Sciences, 80 Nandan Road, Shanghai 200030, China
\and
SHAO--XMU Joint Center for Astrophysics, Xiamen, Fujian 361005, China
\and
University of Chinese Academy of Sciences, 19A Yuquan Road, Beijing 100049, China
}

\date{Received September 30, 20XX}
\abstract
{Type-C quasi-periodic oscillations (QPOs) in black hole X-ray binaries serve as sensitive probes of accretion geometry and variability, although their full evolution throughout the entire outburst and the physics driving the evolution are not yet fully understood.}
{Utilizing the intensive, broadband observations of \textit{Insight}-HXMT on MAXI J1348$-$630, we aim to comprehensively investigate the properties and evolution of type-C QPOs during both its 2019 main and subsequent mini outbursts.} 
{We performed a comprehensive timing and spectral analysis of \textit{Insight}-HXMT data. The evolution of QPO frequency, fractional rms, and energy dependence was tracked across spectral states, and correlations with X-ray flux and spectral shape were examined.}
{(1) The type‑C QPO frequency rose from 0.24 Hz to 7.28 Hz during the main outburst rise and subsequently exhibited a nearly constant value of $\sim$7 Hz when the QPO reappeared across different spectral states. (2) The fractional rms spectrum underwent a pronounced hardening following the transition from the hard to the hard-intermediate state. (3) The correlation between QPO frequency and X-ray flux exhibited strong hysteresis between the rise and decay phases; notably, the hysteresis loop reversed direction between the main and mini outbursts. (4) In contrast, the frequency and hardness followed a tight, nearly single-track anti-correlation.}
{Our results provide a complete picture of type-C QPO evolution. The stable reappearance frequency at $\sim$7 Hz indicates that the Compton region may reform at a consistent characteristic scale across different spectral states. Results (3)–(4) suggest that the type‑C QPO frequency evolution is governed more by spectral shape rather than by X-ray luminosity, while the reversed hysteresis provides a new perspective on the differences between the main and mini outbursts, possibly originating from variations in the initial conditions.}

\keywords{Black hole physics -- X-rays: binaries -- stars: individual: MAXI J1348$-$630}

\maketitle
\nolinenumbers 

\section{Introduction}
\label{sec:intro}

During an outburst, black hole X-ray binaries (BHXRBs) typically exhibit changes in luminosity and X-ray spectra that correspond to transitions between distinct spectral states. In a bright transient outburst, the source usually evolves through the hard state (HS), hard-intermediate state (HIMS), and soft-intermediate state (SIMS) during the rise phase, before entering the soft state (SS). Then it returns to the HS during the decay phase of the outburst \citep{Mendez97, Homan_2005, Remillard_2006, Belloni10, Belloni_2016}. 
Throughout the outburst, the spectral evolution typically shows a hysteresis effect, forming a q-shaped trajectory in the hardness-intensity diagram (HID; e.g., \citealt{Homan_2005, Belloni05, Belloni10}).
Following the bright main outburst, some BHXRBs exhibit much smaller and shorter outbursts, which are called mini-outbursts \citep{Chen_1997, Yan_2017,Zhang_mini_2024}. 

In addition to the X-ray spectral evolutions, BHXRBs also exhibit short-term X-ray variability during the different spectral states. This variability can be studied using the power density spectrum (PDS), which is generated by applying the Fourier transform. The PDS displays the amplitude of variability across different frequencies; it is primarily composed of background noise, but sharp peaks occasionally appear above the noise. These peaks correspond to quasi-periodic oscillations (QPOs) \citep[e.g.,][]{Nowak00, Casella04, Belloni_2014,Ingram20}.
The properties of a QPO are described by three parameters: the central frequency, the quality factor \( Q \) (central frequency divided by the full width at half maximum, FWHM), and the fractional rms amplitude. 
As a normalized measure of variability amplitude, the fractional rms of a QPO quantifies its relative strength against the total mean intensity \citep{VanDerKlis89,Ingram20}.
Based on their central frequency, QPOs are generally divided into two categories: low-frequency QPOs (LFQPOs), which span from a few mHz up to tens of Hz \citep[e.g., ][]{psaltis_1999,Belloni02a, Casella04,Belloni05,Remillard_2006}, and high-frequency QPOs (HFQPOs), which fall in the range of tens to hundreds of Hz \citep[e.g., ][]{Morgan97,Remillard99a,Belloni01,Homan01,Belloni_2014}.

LFQPOs have been extensively studied and are typically classified into three types: A, B, and C \citep{Casella04, Casella05, Motta_2011}. 
Among them, type-C QPOs are the most commonly detected in BHXRBs and can appear in all spectral states \citep[for review, see][]{Ingram20}. 
They typically manifest as prominent and narrow features in the PDS, with a strong fractional rms amplitude of 3--16\% and a high \( Q \) value. 
Although present in all states, type-C QPOs are most frequently observed in the HS and HIMS, accompanied by band-limited noise (BLN) \citep[e.g.,][]{Belloni02a}.
The central frequency of type-C QPOs typically evolves from tens of mHz up to $\sim$10 Hz, and higher frequencies have been detected in some very luminous X-ray sources \citep{Casella04,Belloni05,Motta_2011, Motta_2012,Ingram20}.
Type-B QPOs are typically detected in the SIMS. They have a central frequency of 1--7 Hz and are characterized by a relatively strong fractional rms amplitude and a narrow peak \citep[e.g., ][]{Wijnands99a, Casella04, Casella05,Belloni_2014}.
Recently, the simultaneous presence of type-B and type-C QPOs has been reported in different sources \citep{Pei_2025_simul, wang_2026_co}.
Type-A QPOs are observed primarily in the SIMS and SS as a weak and broad peak, typically having a central frequency in the range of 6--8 Hz and being accompanied by red noise components; they are the least common LFQPOs \citep{Wijnands99a,Homan01,Casella04,Belloni_2014}.

These three typical LFQPOs have been extensively studied and several models have been proposed to explain them. A common explanation is the geometric Lense-Thirring precession of an accretion flow or jet, which arises from the misalignment between the black hole's spin axis and the rotational axis of the inner accretion flow \citep{Stella98, Ingram09, You2018,You_2020,Ma_2023}. Alternatively, models based on intrinsic oscillations and instabilities of the accretion flow have also been used \citep{Debnath_2010,Debnath_2013, Rodriguez2002}. Multiple observational lines of evidence support the geometric model as the most promising explanation, such as the dependence on inclination angle of the QPO amplitude and phase lag \citep{Motta_2015,Heil_2015,vandenEijnden2017,Ma_2021} and the modulation of the iron line equivalent width and centroid energy over the QPO cycle \citep{Ingram_2014,Ingram_2016}. 
Recent works have further supported the association of type-C QPOs with a horizontally extended corona, citing evidence from inclination dependence \citep{Motta_2015, vandenEijnden2017} on the geometric evolution of the accretion flow \citep{Alabarta24, Davidson_2025, Choudhury_2025}.

\begin{figure*}
\centering
\includegraphics[width=0.9\linewidth]{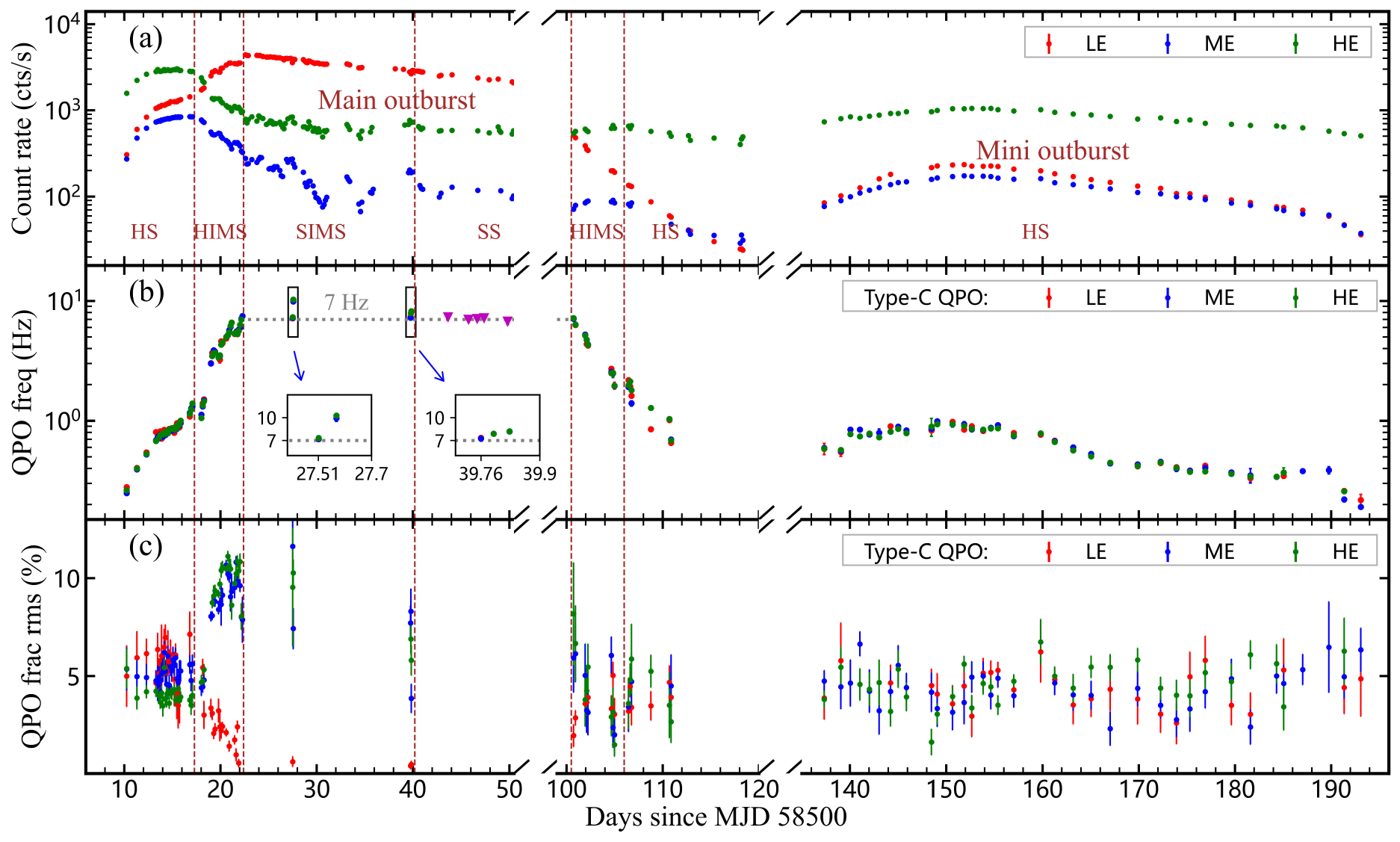}
\caption{Evolution of (a) the count rate, (b) the centroid frequency, and (c) the fractional rms of the type-C QPO as observed with the three \textit{Insight}-HXMT instruments. For comparison, the frequencies of type-A QPOs reported by \cite{zhang_2023} are also plotted in panel (b) (magenta triangles). Vertical brown dashed lines demarcate the different states during the main and mini-outbursts.}
\label{fig:QPO-evolution}
\end{figure*}

\begin{figure*}
\centering
\includegraphics[width=1\linewidth]{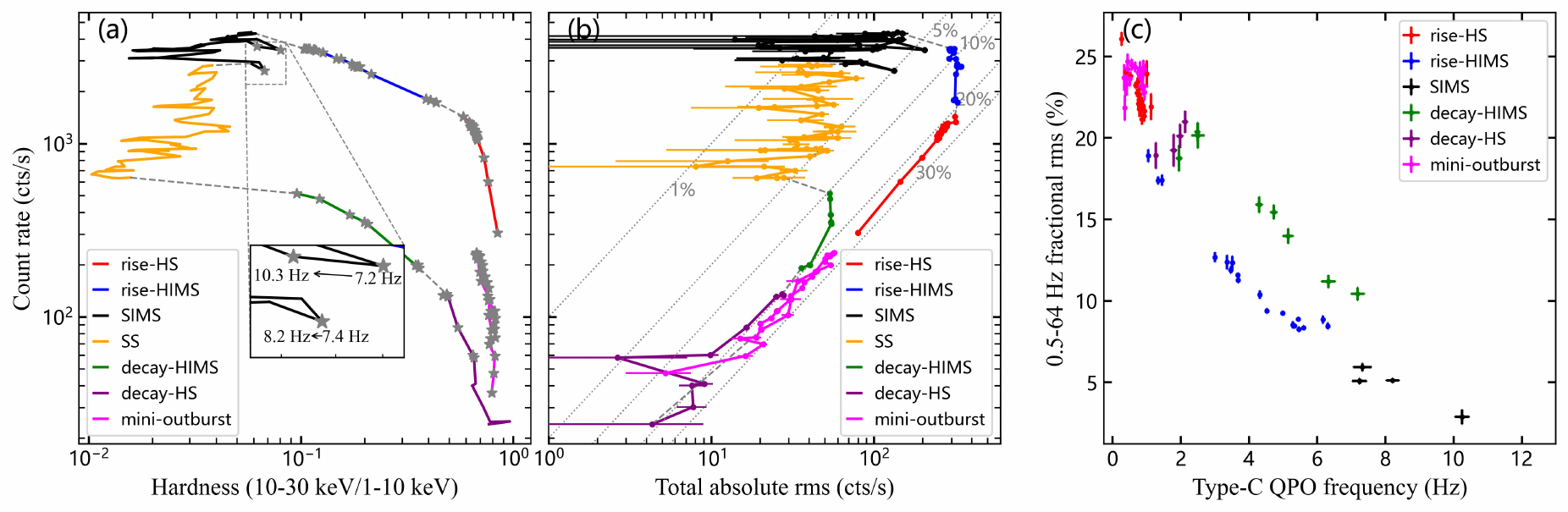}
\caption{(a) HID, where the count rate is measured in the 1--10 keV band and the hardness is defined as the ratio of the 10--30 keV to the 1--10 keV count rates; gray pentagrams mark the detections of type-C QPOs.
(b) RID, where the fractional rms is calculated over the 0.5--64 Hz frequency range in the 1--10 keV band.
(c) Type-C QPO frequency versus the total fractional rms (computed in the 1--10 keV band over 0.5--64 Hz).
In all three panels, different colors represent different spectral states.}
\label{fig:hid-rid}
\end{figure*}

\section{MAXI J1348$-$630}
\label{MAXIJ1348}
MAXI J1348$-$630 was discovered during its 2019 outburst \citep{Yatabe2019}.
\cite{Zhang2020} identified the source as a BHXRB and characterized its distinct outburst states. The source underwent a bright complete outburst lasting approximately 110 days, followed by several mini-outbursts. Hereafter, we refer to this complete outburst as the main outburst, and the term "mini-outburst" specifically refers to the first mini-outburst of MAXI J1348$-$630.

All three typical LFQPOs (A, B, and C) have been detected \citep{Zhang2020}.  
Among them, type-B QPOs have been suggested to be connected to jet activity by several studies \citep{Belloni2020,Zhang2021,Liu_2022,Carotenuto2022MNRAS}, while type-A QPOs are suggested to originate in an extended Comptonization region \citep{zhang_2023}.  
In addition, spectral–timing modeling of the type-B QPO indicates that two Comptonization regions are required to reproduce its observed properties \citep{Garc2021,Bellavita_2022}.
As the most common type of LFQPO, type-C QPOs in MAXI J1348$-$630 exhibit a characteristic frequency evolution spanning from 0.3 Hz to 2.9 Hz in \textit{NICER} and \textit{AstroSat} data, along with a typical rms spectrum that rises at low energies and flattens (or slightly declines) at high energies, where the flattening energy shifts from 2 to 5 keV across different phases \citep{Alabarta_2022, Jithesh2021, Bhowmick_2022}.
Furthermore, type-C QPOs exhibit hard phase lags and remain consistent across different outburst states, supporting a Comptonization origin of them \citep{Alabarta_2022}. By tracking the evolution of the Comptonization region, \cite{Alabarta24} further attributes the origin of the type-C QPO to a horizontally extended corona.

Although previous studies have significantly explored type-C QPOs in this source, time gaps in their data from telescopes like \textit{NICER} resulted in a lack of coverage of several key stages of the QPO evolution, such as the HIMS, SIMS, and early decay phase.
The intensive observations of \textit{Insight}-HXMT allow us to perform a more detailed and comprehensive analysis of type-C QPOs across a much broader X-ray energy band. We describe the observation and data reduction in \autoref{sec:data}, present the timing analysis and results in \autoref{sec:results}, discuss our findings in \autoref{sec:discussion}, and conclude in \autoref{sec:conclusion}.

\section{Observation and data analysis}
\label{sec:data}
\textit{Insight}-HXMT, an X-ray astronomical satellite launched on June 15, 2017, features a broad energy coverage spanning 1 to 250 keV.
Its three primary payloads include the Low Energy X-ray Telescope (LE, 1--15 keV), the Medium Energy X-ray Telescope (ME, 5--30 keV), and the High Energy X-ray Telescope (HE, 20--250 keV) \citep{zhang_overview_2020}. Hereafter, the energy bands of 1--10 keV, 10--30 keV, and 27--200 keV are denoted as "LE" (low energy), "ME" (middle energy), and "HE" (high energy), respectively.
Each \textit{Insight}-HXMT observation comprises multiple data segments termed "exposures", where each segment is uniquely identified by an Exposure ID. 

We used the \textit{Insight}-HXMT Data Analysis Software package (\texttt{HXMTDAS}) V2.05 to process the raw data.
By setting the time resolution to 1/256 seconds, we obtained both the raw and the background light curves. 
Using the \texttt{powspec} tool, we generated the average PDSs from the raw light curves with a segment duration of 64-s, corresponding to a frequency resolution of 0.0156 Hz. 
Since type-C QPOs exhibit relatively high frequencies during the HIMS and SIMS ($\sim$1--10 Hz), we adopted a shorter segment duration of 8-s in place of 64-s in these states to increase the number of segments for averaging and thus improve the signal-to-noise ratio (SNR) of the PDSs.
Furthermore, during the mini-outburst, when the photon count rates are low and the PDSs evolve slowly, we merged exposures within a single observation to obtain more segments for averaging.
The resulting PDSs were logarithmically rebinned with a geometric binning factor of 1.03.

All PDSs from the \textit{Insight}-HXMT were normalized using Leahy normalization \citep{Leahy1983}, and the Poisson noise was subtracted. We formatted the PDSs for compatibility with \texttt{XSPEC} (version 12.10.1) using the \texttt{flx2xsp} tool \citep{Ingram12}, allowing us to use \texttt{XSPEC} models for data fitting.
We fitted the PDSs using multiple \texttt{lorentz} components in \texttt{XSPEC} and employed chi-square statistics to evaluate the goodness-of-fit. For exposures with low SNR where the type-C QPO was poorly constrained, we fixed its \( Q \) at 8 to facilitate a stable fit by using the \texttt{newpar} command to define the FWHM as 0.125 times the centroid frequency. The best-fitting parameters and their 1$\sigma$ uncertainties were derived from the chains generated by the Monte Carlo Markov Chain (MCMC) algorithm \texttt{emcee}, implemented in \texttt{XSPEC}\footnote{\url{https://github.com/zoghbi-a/xspec_emcee}}.
The fractional rms amplitude of the QPO was computed as 
$\sqrt{\texttt{norm}/\langle C_{\rm raw} \rangle}$ 
\citep{vanderKlis1988,VanDerKlis89}, 
where $\texttt{norm}$ is the integrated Leahy-normalized power of the QPO \texttt{lorentz} component and $\langle C_{\rm raw} \rangle$ is the mean count rate of the raw light curve. 
The rms values were then corrected for background dilution by multiplying a factor of $(S+B)/S$, where $S$ and $B$ denote the source and background count rates, respectively.

\begin{figure*}
\centering
\includegraphics[width=1\linewidth]{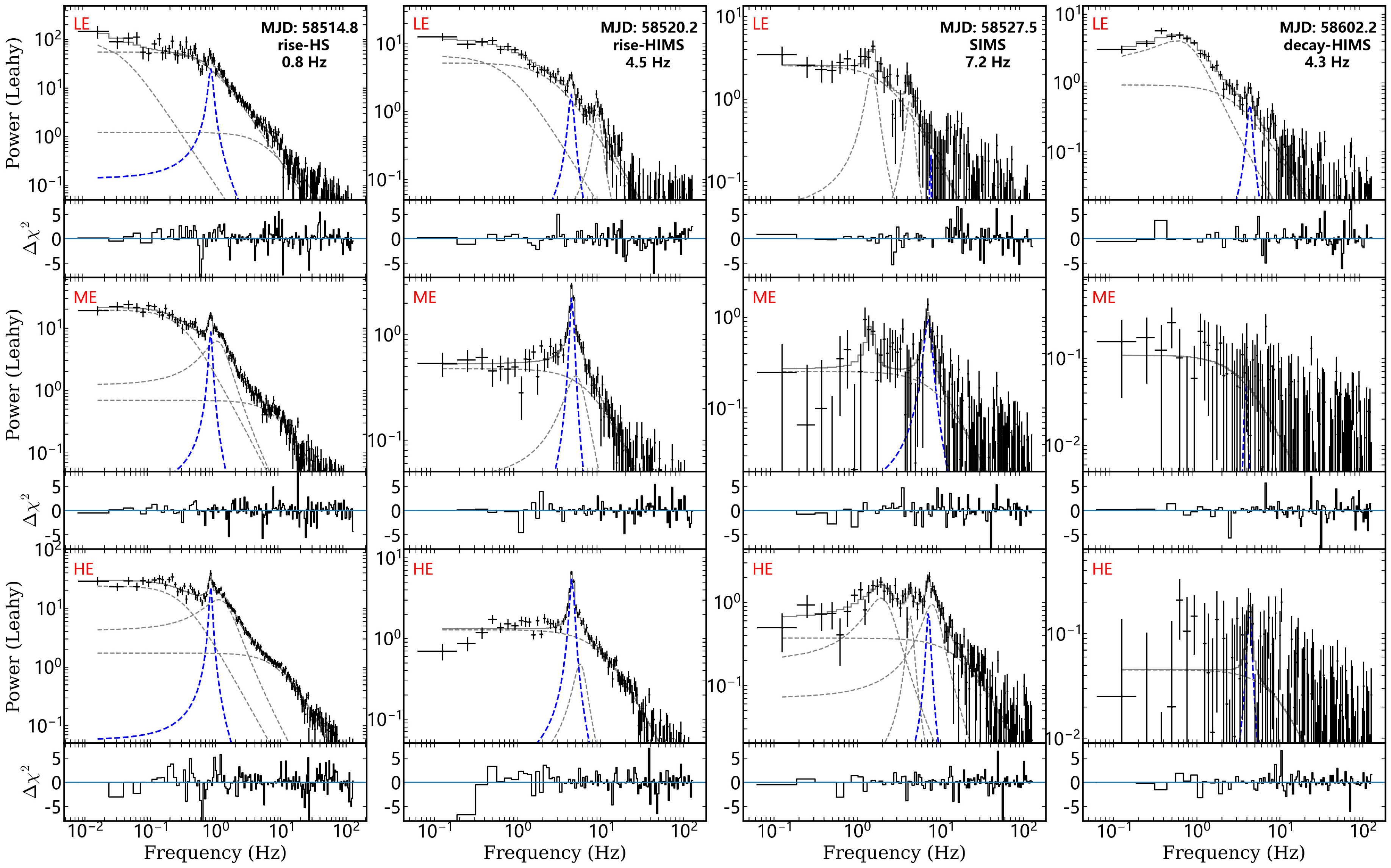}
\caption{Representative PDSs in the LE, ME, and HE bands during different outburst states. The blue dashed lines indicate the fitted type-C QPO components.}
\label{fig:pds}
\end{figure*}

\section{Results}
\label{sec:results}

We present an overview of the count rate evolution across different spectral states in panel (a) of \autoref{fig:QPO-evolution}. 
The classification of states is based on the HID and rms–intensity diagram (RID, \citealt{munoz2011}) shown in \autoref{fig:hid-rid}, and is broadly consistent with the state identifications in \citet{Zhang2020} and \citet{Carotenuto_2025}.
In the following, we use “rise-HS” and “rise-HIMS” to denote the HS and HIMS during the rise phase of the main outburst, and “decay-HS” and “decay-HIMS” for those during the decay phase.

Type-C QPOs are detected across the HS, HIMS, and SIMS; their identification is supported by the correlation between the QPO frequency and the broadband fractional rms shown in panel (c) of \autoref{fig:hid-rid}, as commonly used for QPO classification \citep{Casella05, Alabarta_2022}.
In \autoref{fig:pds}, we present representative PDSs in three energy bands from different outburst states. These PDSs show that the type-C QPOs evolve in frequency across epochs and are consistently accompanied by typical BLN components. In the following, we present the detailed evolution of the type-C QPO properties.

\subsection{Frequency evolution of type-C QPO}  
\label{sec:freq_evolution}  
As shown in panel (b) of \autoref{fig:QPO-evolution}, the frequency of the type-C QPO exhibits a clear evolution with the outburst state.
In the main outburst, the frequency increased from $\sim$0.24 Hz to $\sim$7.28 Hz during the rise-HS and rise-HIMS. As the source transitioned to the SIMS, the type-C QPO was intermittently detected \citep{Liu_2022}, with its frequency observed within a narrow range of 7.16--10.25 Hz. After approximately two months in the SS, the type-C QPO reappeared during the decay phase, with its frequency decreasing from $\sim$7.24 Hz to $\sim$0.64 Hz. This frequency evolutionary behavior is typical of type-C QPOs in BHXRBs \citep[e.g.,][]{Casella04, Debnath_2013,Ingram20,Zhangyuexin_2024}. During the mini-outburst, the source remained in the HS. The QPO frequency first increased to $\sim$0.99 Hz around the X-ray flux peak and then decreased to $\sim$0.20 Hz. This type-C QPO evolution observed during a mini-outburst has also been reported in other sources \citep{wang_failed_2022}, but such cases are rare, likely due to the low X-ray intensity.

Additionally, the intensive detection of type-C QPOs in MAXI J1348$-$630 reveals a new detail in their frequency evolution. During the main outburst, as the outburst reached the end of the rise-HIMS at MJD 58522.3, the frequency rose to a peak of $7.28^{+0.06}_{-0.07}$ Hz. Then the outburst entered the SIMS and the type-C QPO temporarily disappeared. After a few days, the type-C QPO reappeared twice at MJD 58527.51 and 58539.76, respectively. Interestingly, its frequency remained nearly consistent with its pre-disappearance value, as shown in the insets of \autoref{fig:QPO-evolution}. Specifically, the frequencies measured in these reappearances are $7.16^{+0.09}_{-0.10}$ Hz and $7.36^{+0.30}_{-0.31}$ Hz, respectively. Furthermore, when the type-C QPO reappeared at the beginning of the decay phase, it again exhibited a similar frequency of $7.24^{+0.23}_{-0.29}$ Hz. Despite these four detections occurring in different spectral states that spanned approximately 80 days, they demonstrate a stability of frequency, with a mean frequency of 7.26 Hz and a standard deviation of 0.09. This indicates that the type-C QPO displays a stable characteristic frequency from the end of the rise-HIMS to the start of the decay phase when the QPO reappeared occasionally. In contrast, the count rate dropped by a factor of several during the same period.

\begin{figure*}
\centering
\includegraphics[width=1\linewidth]{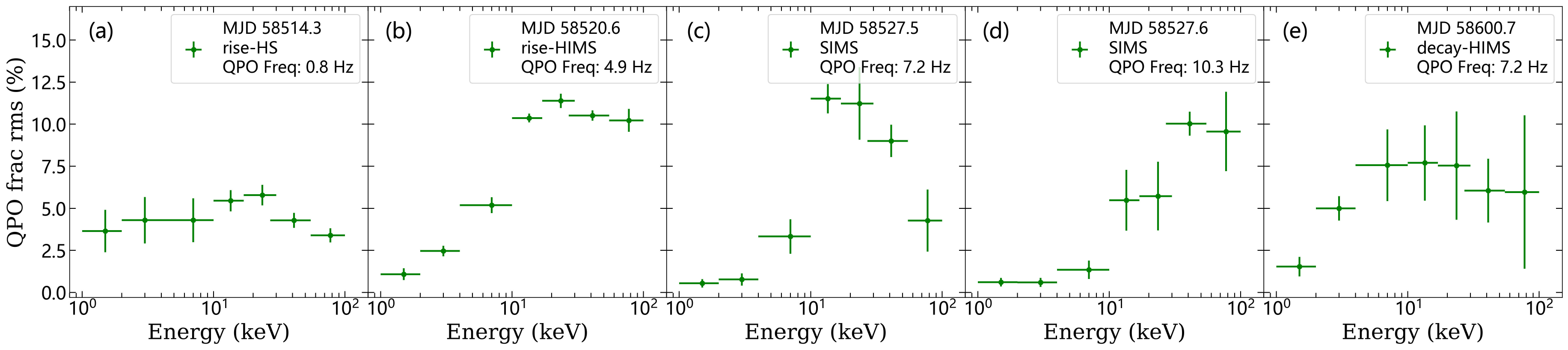}
\caption{Evolution of the fractional rms spectrum of type-C QPO. Each panel is labeled in the upper-right corner with the observation date, spectral state, and type-C QPO frequency.}
\label{fig:CQPO_rms_spectrum}
\end{figure*}

In addition to the stable frequency spans tens of days, the frequency of type-C QPO exhibits rapid increases on short timescales during the SIMS \citep[see also][]{Liu_2022}. As shown in the insets in panel (b) of \autoref{fig:QPO-evolution}, on MJD 58527.51, the frequency increased from 7.16 Hz to 10.25 Hz over approximately 6000 seconds. Similarly, around MJD 58539.76, it increased from 7.36 Hz to 8.03 Hz within 2000 seconds, and then to 8.19 Hz after another 2500 seconds. We used the HID to explore how these rapid frequency increases relate to changes in the spectral hardness. 
As shown in \autoref{fig:hid-rid}, the appearance of a type-C QPO coincided with the highest hardness level of SIMS. From panel (b) of \autoref{fig:hid-rid}, it can also be seen that some of these type-C QPO detections are associated with broadband rms values exceeding 5\%. This suggests that, during these short intervals, the source may have undergone brief excursions toward a relatively harder and noisier configuration, closer to the HIMS \citep[see also][]{Liu_2022}. Following these harder intervals, the QPO frequency increased rapidly as the spectrum softened.
Overall, this pattern suggests an evolutionary scenario in the SIMS where the type-C QPO initially reappeared in a relatively hard spectrum, which then evolved into a softening spectrum accompanied by a rapid increase in QPO frequency.

\subsection{Fractional rms evolution of type-C QPO}  
\label{sec:rms_evolution}  
As shown in panel (c) of \autoref{fig:QPO-evolution}, during the rise-HS, the fractional rms exhibited comparable amplitudes across the LE, ME, and HE bands, with the LE band generally being higher than the HE band. However, after the source transitioned to the rise-HIMS, the QPO rms in the ME and HE bands increased, while those in the LE band decreased, resulting in substantially higher rms in the ME and HE bands compared to the LE band. 

To obtain a more detailed energy-dependent view of the fractional rms evolution, we analyzed the type-C QPO across subdivided energy bands. The energy bands were defined as follows: 1--2 keV, 2--4 keV, and 4--10 keV (LE); 10--17 keV and 17--30 keV (ME); 27--55 keV and 55--100 keV (HE). \autoref{fig:CQPO_rms_spectrum} presents fractional rms spectra of the type-C QPO from five periods spanning from the rise-HS to the decay-HIMS. For rise-HS, rise-HIMS, and decay-HIMS, we selected one exposure with a high photon count for each phase; for SIMS, we selected two intervals characterized by a rapid increase in type-C QPO frequency (from 7.2 to 10.3 Hz) to analyze its effect on the fractional rms spectrum. 

During the HS, the fractional rms of the type-C QPO varies mildly across energies, ranging between 3\%–6\%. It slightly increases at low energies, followed by a gradual decline at higher energies (see panel (a) of \autoref{fig:CQPO_rms_spectrum}). This trend is broadly consistent with the results of \cite{Alabarta_2022} (using \textit{NICER} data) and \cite{Jithesh2021} (using combined \textit{NICER} and \textit{AstroSat} data).
In the following rise-HIMS, the fractional rms spectrum hardens significantly. The rms amplitude exceeds 10\% in all energy bands above 10 keV, while it decreases in the lowest two energy bands (1–2 keV and 2–4 keV; see panel (b) of \autoref{fig:CQPO_rms_spectrum}).

For the next two SIMS intervals (panels (c) and (d) of \autoref{fig:CQPO_rms_spectrum}), the fractional rms spectra are generally similar to those in the rise-HIMS, exhibiting high amplitudes at high energies. However, the peak in panel (d) shifts to higher energies than in panel (c), pointing to a correlation whereby a higher QPO frequency corresponds to a harder rms spectrum. Finally, in the decay-HIMS (panel (e)), although the type-C QPO is at relatively high frequency, the rms spectrum returns to a softer profile, peaking at $\sim$10 keV. Together, \autoref{fig:QPO-evolution} and \autoref{fig:CQPO_rms_spectrum} reveal that the rms spectrum of the type-C QPO undergoes remarkable changes across different states, characterized most notably by its hardening during the rise-HIMS and SIMS.

\subsection{Correlation between type-C QPO and X-ray flux}
\label{sec:fre-spectra}
\begin{figure*}
\centering
\includegraphics[width=1\linewidth]{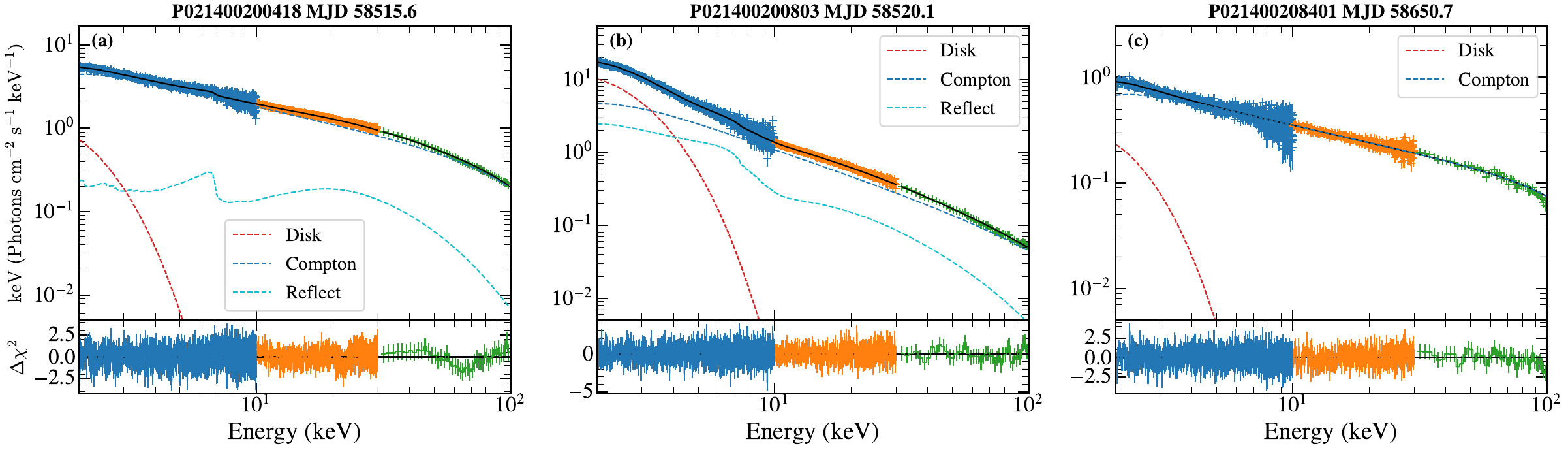}
\caption{Representative spectra observed in the rise-HS (panel a), rise-HIMS (panel b) and mini-outburst (panel c). The data points with error bars correspond to the \textit{Insight}-HXMT LE (blue), ME (orange), and HE (green) observations. The solid black line represents the best-fitting total model, while the dashed lines indicate the individual components: the accretion disk (red), Comptonization (blue), and reflection (cyan), respectively.}
\label{fig:spectra_fit}
\end{figure*}

We further investigate how the type-C QPO frequency generally evolves with X-ray flux and hardness. To derive the X-ray flux, we performed spectral fitting using XSPEC v12.12.1 with the \texttt{chi} statistic. For the spectra in the rise-HS, rise-HIMS, and SIMS, we used the model \texttt{constant*TBabs*(diskbb + nthcomp + relxillCp*nthratio) \citep[see also ][]{Wu_2025}.}
The \texttt{constant} is used to cross-calibrate the three instruments in the joint fits, with the LE value fixed to 1. The Comptonization emission from the corona was modeled using \texttt{nthcomp}, with its seed photon temperature $kT_{\mathrm{bb}}$ tied to the disk temperature $T_{\mathrm{in}}$. The photon index $\Gamma$ and the electron temperature $kT_e$ in \texttt{relxillCp} were tied to those in \texttt{nthcomp}, and the reflection fraction was fixed at $-1$ to separate the Comptonized continuum from the reflection component. The \texttt{nthratio}\footnote{\url{https://github.com/garciafederico/nthratio}} is included to account for the soft excess caused by fixing the seed photon temperature ($kT_{\rm bb}=0.01$ keV) in \texttt{relxillCp}. A prescribed density of the hydrogen column ($N_H$) $0.86 \times 10^{22}$ cm$^{-2}$ was applied in the \texttt{TBabs} \citep{Carotenuto2021MNRAS}. We adopted the fixed black hole spin a = 0.79 \citep{Guan2024}, the inclination angle $i=29.3^{\circ} $\citep{Carotenuto2022MNRAS}, the black hole mass 11$M_{\odot }$ \citep{Lamer2021} and the distance $D=3.39$ kpc \citep{Lamer2021}. A continuous power-law emissivity profile was adopted for simplicity in \texttt{relxillCp}, with it set to $q_{1} = q_{2} = 3.0$. In the decay-HIMS, decay-HS, and the mini-outburst, the spectra were well described by \texttt{diskbb} plus \texttt{nthcomp}, without requiring a reflection component \citep{Kumar2022,Dai23}. \autoref{fig:spectra_fit} displays examples of the spectral fits. Then the \texttt{cflux} was used to calculate the flux of different components in the 0.1–100 keV energy band.

\begin{figure*}
\centering
\includegraphics[width=0.8\linewidth]{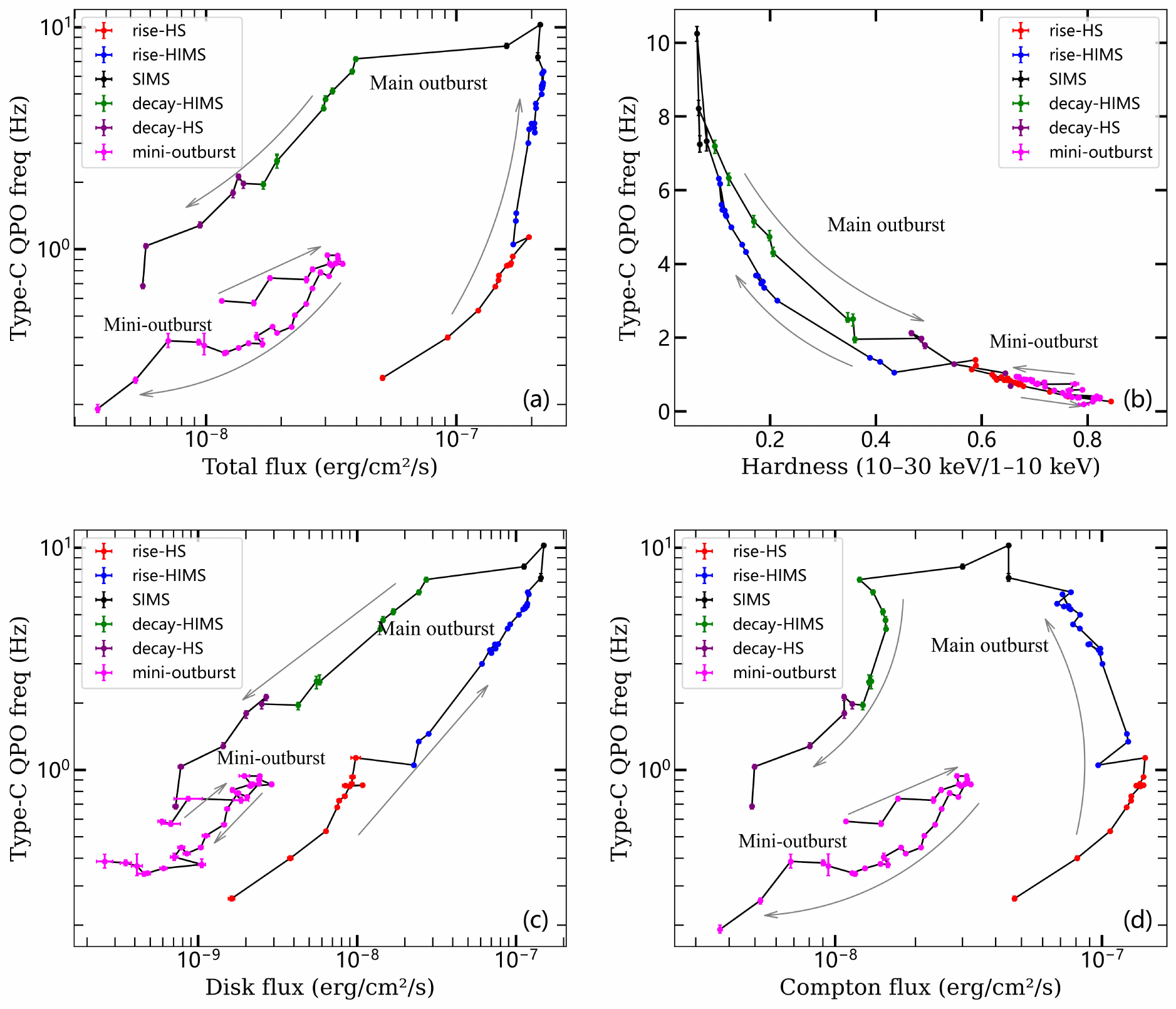}
\caption{Correlations of the type-C QPO frequency with: (a) total X-ray flux (0.1--100 keV), (b) spectral hardness, (c) disk flux, and (d) Comptonization flux. Evolutionary directions are indicated by arrows.}
\label{fig:freq-relationship}
\end{figure*}

Panel (a) of \autoref{fig:freq-relationship} shows the correlation between the type-C QPO frequency and the total X-ray flux. 
The frequency generally exhibits a positive correlation with the total flux.
In the main outburst, a clear hysteresis loop can be observed in the flux-frequency correlation, with the flux at a given QPO frequency being several times higher in the rise phase than in the decay phase. A similar flux difference also emerges in the mini-outburst, but interestingly, the behavior is reversed: the decay phase flux is higher than that of the rise phase at the same frequency. Thereby, as the arrows indicate, the hysteresis loops in the main and mini-outbursts evolve in opposite directions, with the main outburst tracing a counterclockwise loop and the mini-outburst a clockwise one. 
Similar flux–frequency correlations and hysteresis effects of type-C QPOs have been reported previously \citep[e.g.,][]{Heil_2011,Motta_2011}; in our case, the extensive QPO coverage provided by $\textit{Insight}$-HXMT allows us to trace the continuous evolution path throughout both the main and mini-outbursts in remarkable detail.

Panel (b) of \autoref{fig:freq-relationship} displays a negative correlation between type-C QPO frequency and hardness. Compared to the flux-frequency relation in panel (a), this hardness-frequency relation follows a nearly single, much tighter trend across different outburst phases, although a slight distinction is still discernible between the rise-HIMS and the decay-HIMS. This demonstrates a robust anti-correlation that is largely independent of the outburst evolution. A similar anti-correlation between the type-C QPO frequency and hardness has also been reported in other sources (e.g. \citealt{Stiele2013,Mendez_2022,Garc_2022_mn,zhu_2024_1727}).

Panels (c) and (d) of \autoref{fig:freq-relationship} show the correlations between the QPO frequency and the flux from the individual spectral components, i.e., the disk and Comptonization, respectively. The fewer data points at the lowest frequencies for the mini-outburst in panel (c) are due to the Comptonization-dominated emission, as a disk component was not required in the spectral fit for these observations.
The QPO frequency exhibits a positive correlation with the disk flux, which is similar to that of total flux but follows a more consistent track across different outburst phases. Meanwhile, panel (d) reveals a more complex relationship with the Comptonization flux. During the HS, the frequency increases with the Comptonization flux. However, upon transition into the HIMS, the QPO frequency continues to rise while the Comptonization flux declines, leading to an anti-correlation. This anti-correlation is also observed during the decay-HIMS, until it eventually reverts to a positive correlation as the type-C QPO frequency continues to decrease. This complex behavior with the Comptonization flux may indicate a connection between the type-C QPO and the evolution of the Comptonizing corona, as also suggested by spectral–timing studies using variable-Comptonization models (e.g. \citealt{Garc_2022_mn,Zhang_yuexin_2022,Rawat_2023}).

\section{Discussion}
\label{sec:discussion}
As presented in \autoref{sec:results}, the extensive detections of type-C QPOs in MAXI J1348$-$630 unveil their properties in great detail. The key findings can be summarized as follows:
\begin{itemize}
    \item The type-C QPO exhibited a stable reappearance frequency of $\sim$7 Hz from the end of the rise-HIMS to the start of the decay phase.
    \item After the transition from the rise-HS to the rise-HIMS, the fractional rms spectra of the type-C QPO hardened significantly.
    \item The flux-frequency correlation exhibits a hysteresis loop that evolves in opposite directions in the main and mini-outbursts, whereas the QPO frequency and hardness follow a single, tight anti-correlation throughout both main and mini-outbursts.
\end{itemize}

These findings demonstrate the long-term evolution of type-C QPOs across different outburst phases. In the following subsections, we will discuss these results and their underlying physics.

\subsection{The stable reappearance frequency at $\sim$7 Hz}
\label{sec:discuss-stable-freq}
The frequency of type-C QPO typically evolves by increasing with the X-ray intensity during the outburst rise and decreasing during the outburst decay \citep[e.g.,][]{Chakrabarti_2008,Debnath_2013,Zhangyuexin_2024}. 
During the SIMS, type-C QPO is generally observed to disappear \citep{Motta16,Ingram20}, except for rare detections of short-lived reappearance \citep{Stiele_2023,Li_2025}. Consequently, the evolution of type-C QPO during this period remains highly uncertain.
Thanks to the intensive observations by \textit{Insight}-HXMT, our results show that the type-C QPO exhibits a stable 7 Hz frequency when it reappears during the SIMS, a value consistent with that at the end of the rise-HIMS and at the beginning of the decay-HIMS (see \autoref{fig:QPO-evolution}).
Interestingly, type-A QPOs detected during the SS also exhibit a similar stable frequency of 7 Hz \citep{zhang_2023}, as marked by the magenta inverted triangles in panel (b) of \autoref{fig:QPO-evolution}. This coincidence in frequency not only suggests a possible connection between the two types of QPOs but also highlights the special significance of the 7 Hz value, which likely corresponds to a stable structure or component across different spectral states. The phenomenon that type-C and type-A QPOs exhibit a similar frequency has also been reported in GX 339$-$4 \citep{Motta_2011} and Swift J1658.2$-$4242 \citep{Bogensberger_2020}, implying that the connection between type-C and type-A QPOs might be more common in BHXRBs, though it could be rarely detected due to the lack of intensive monitoring observations.

Type-C QPOs have been extensively linked to the Comptonization component in both spectral and timing studies \citep[e.g.,][]{Stiele2013, Gao_2023, Choudhury_2025}. In MAXI J1348$-$630, this interpretation is further supported by multiple investigations, particularly through the rms and phase-lag spectra of type-C QPOs and their successful modeling with Comptonization corona \citep{Zhang2020,Alabarta_2022,Alabarta24}. Our analysis also reveals that the reappearance of type-C QPOs during the SIMS coincides with spectral hardening, characterized by higher hardness ratios in the HID (\autoref{fig:hid-rid}) and a lower photon index \citep[see also][]{Liu_2022}, providing additional evidence for their coronal (Comptonization) origin. Similarly, type-A QPOs are likely to arise from the Comptonization component, as their QPO spectra closely follow the shape of the Comptonized continuum \citep{zhang_2023}. The comparable characteristics of type-C and type-A QPOs, such as their connection to Comptonization, similar centroid frequencies, and consistent flux–frequency correlations \citep{Motta_2011}, suggest that both phenomena may originate from a common Comptonization region that persistently reappears across SIMS and SS.

If the type-C QPO frequency represents a characteristic scale, the persistent reappearance of the $\sim$7 Hz frequency implies that the originating Comptonization region, which was previously cooled, diluted, or geometrically disrupted, has been reformed or refilled on a consistent scale. This reformation appears to be independent of the X-ray luminosity, although the physical origin of this specific scale remains unclear. Furthermore, this characteristic scale is not the minimum possible one, as evidenced by the rapid frequency increasing to $\sim 10$ Hz during the SIMS \citep[see \autoref{fig:QPO-evolution} and also][]{Liu_2022}. Together with the associated spectral softening (\autoref{fig:hid-rid}), these observations point to a coherent evolutionary picture for the Comptonization region. This region reforms at a characteristic scale during the SIMS and even the SS, then experiences rapid cooling that triggers its contraction and faster oscillation, leading to the observed increase of QPO frequency. 

Although current type-C QPO models generally invoke the Comptonization region as their origin, they do not account for its reappearance during the SIMS. Therefore, the stable reappearance frequency of the type-C QPO places special constraints on these theoretical models. For instance, in the propagating oscillatory shock (POS) model \citep{Chakrabarti_2008,Debnath_2010,Debnath_2013}, it requires the shock to maintain a constant location when it reappears during the SIMS, despite the substantial decline in accretion rate; in the Lense-Thirring precession model of a hot inner flow \citep{Ingram09,Ingram12,Ingram_2016}, the stable frequency requires that the outer radius of the flow be consistently reestablished at the same characteristic scale. These requirements pose a significant implication and even challenge to the models, and their precise underlying mechanisms require further investigation.

\subsection{The hardening of the fractional rms spectra}
\label{sec:discuss-rms}
In BHXRBs, the fractional rms of type-C QPOs exhibits a significant energy dependence. The rms typically decreases with energy below $\sim$2 keV, which is attributed to dilution by the thermal disk emission with little variability \citep{Casella04}. At higher energies, the fractional rms spectra of type-C QPOs commonly show a flattening above a certain energy. This high-energy flattening has been observed in sources such as XTE J1859+226 \citep{Casella04}, MAXI J1631$-$479 \citep{Bu_2021}, MAXI J1803$-$298 \citep{Chand_2022}, and MAXI J1535$-$571 \citep{Huang_2018}.
Additionally, evolution of the fractional rms spectra is observed in some sources. For example, in GX 339–4, the rms spectrum hardened as the QPO frequency exceeded a threshold of 1.7 Hz \citep{Zhangliang_2017}; in MAXI J1820+070, the break energy at which the rms spectrum flattened shifted from $\sim$2 keV in the HS to $\sim$10 keV in the HIMS \citep{Ma_2023}. These complex rms spectra have been understood as resulting from different radiative processes altering the relative contribution of photons carrying the QPO variability to the total emission.

Although the type-C QPO has a Comptonization origin, this does not imply that all Comptonized photons carry the QPO variability. Therefore, the observed rms amplitude could be diluted by other, a non-QPO Comptonization components. Within this framework, a possible explanation for the evolution of the rms spectrum in MAXI J1348$-$630 is as follows: in the HS, the QPO's rms is diluted at high energies by the other strong Comptonization component related to the compact jet; this dilution effect decreases in the HIMS/SIMS, which unveils the hard rms spectrum of the QPO from the hot inner flow or a horizontally extended corona. This scenario is supported not only by the jet evolution reported in the radio analysis of \cite{Carotenuto2021MNRAS}, but also by the observed geometric evolution of the accretion system in MAXI J1348$-$630, where the corona is found to be horizontally extended before the SIMS \citep[e.g.,][]{Alabarta24, Davidson_2025}.

\subsection{Hysteresis loops in flux-frequency diagram}
\label{sec:discuss-flux-freq}
The frequency of type-C QPOs shows a positive correlation with X-ray flux across different outburst phases. However, they do not follow a single positive correlation. In the main outburst, for the same QPO frequency, the flux during the decay phase can be up to seven times lower than during the rise phase. Such hysteresis phenomena in the flux-frequency correlation have been reported \citep{Heil_2011,Motta_2011, Debnath_2013,Kubota_2024}.
Compared to the flux-frequency relation, hysteresis in the HID has been more widely observed and studied. However, given the tight correspondence between QPO frequency and hardness as shown in \autoref{fig:freq-relationship}, the two hysteresis phenomena might share a common physical origin.

Although the underlying mechanism of the hysteresis effect remains unclear, it may be related to differences in the magnetic field configuration between the rise and decay phases \citep{Petrucci_08, Begelman_2014}. For instance, \citet{Petrucci_08} proposed the transitions between accretion flow—such as from a disk to a hot inner flow, or the reverse—occur at different magnetic field during the rise and decay. This asymmetry in the transition threshold leads to different accretion structures at similar accretion rates between the rise and decay phases, ultimately producing the hysteresis loop seen in the HID. Naturally, these structural differences would also affect the scale of the Comptonizing region responsible for the type-C QPO, thereby reproducing the similar hysteresis behavior in the flux-frequency relation. 

Furthermore, the hysteresis loop in the mini-outburst evolves in the opposite direction to that in the main outburst. This implies that, although the mini-outburst exhibits spectral and HID properties similar to those of the main outburst \citep{Yan_2017,Bhowmick_2022, Dai23}, the two outbursts may have possessed different initial conditions. As indicated by radio evolution in \cite{Carotenuto2021MNRAS}, in the mini-outburst, the magnetic field is weaker during its rise phase compared to the decay phase, which is opposite to the configuration in the main outburst. This reversed magnetic field evolution could naturally explain the opposite hysteresis trajectories, although the origin of this magnetic field difference requires further investigation.

\section{Conclusion}
\label{sec:conclusion}
We have presented a full evolution of type-C QPOs in the black hole binary MAXI J1348–630 using broadband data from \textit{Insight}-HXMT. The high-cadence detections allowed us to track QPO properties across both the main outburst and the mini-outburst, leading to the following key physical insights. The type-C QPO consistently reappeared at $\sim$7 Hz over a long duration, a value similar to that of the type-A QPO in the SS. This stability suggests that the Comptonizing region reforms at a specific geometric scale, largely independent of the instantaneous accretion rate. After the transition from the HS to the HIMS, the significant hardening of the fractional rms spectrum implies a growing contribution from high-energy photons of non-QPO Comptonization component, possibly linked to weakening in compact jet activity.
Furthermore, we observed a tight, single-track correlation between QPO frequency and spectral hardness, whereas the correlation with X-ray flux exhibited significant hysteresis. This indicates that the QPO mechanism is more directly governed by the spectral shape than by the X-ray luminosity. We also find that the reversal of the flux-frequency hysteresis direction between the main outburst and the mini-outburst, potentially arising from differences in their initial conditions, possibly in the magnetic field configuration of the accretion flow.

\section{Data availability} 
The data underlying this article are available in the \textit{Insight}-HXMT public archive.

\begin{acknowledgements}
This work made use of the publicly available data and software from the \textit{Insight}-HXMT mission. The \textit{Insight}-HXMT project is funded by the China National Space Administration (CNSA) and the Chinese Academy of Sciences (CAS).
X.L.W. \& R.Y.M. were supported by the National Key R\&D Program of China (Grant no. 2023YFA1607902).
F.G.X. \& R.Y.M. were supported by the National SKA Program of China (no. 2020SKA0110102).
J.F.W. was supported by the National Key R\&D Program of China (Grant no. 2023YFA1607904).
This work was supported in part by the Natural Science Foundation of China (NSFC, grants 12373049, 12361131579, 12373017, 12192220, 12192223, 12033004, 12221003, U2038108 and 12133008).
\end{acknowledgements}

\bibliographystyle{aa}
\bibliography{ms}

\end{document}